\documentclass[letterpaper]{article}
\usepackage{osameet3}
\usepackage{amsmath,amssymb,dsfont,color,float,empheq}
\usepackage{graphicx,wrapfig,multicol,multirow}
\usepackage{psfrag,setspace,wrapfig,subfig}
\usepackage[font=footnotesize]{caption}

\usepackage{tikz,pgfplots}
\usetikzlibrary{positioning,shadows,shapes,backgrounds,calc}

\definecolor{orange}{rgb}{1,0.7,0}

\newcommand{\tnr}[1]{{\textnormal{#1}}}

\newcommand{\dij}[0]{d_{ij}}

\newcommand{\dip}[0]{d_{ip}}

\newcommand{\set}[1]{\{#1\}}

\newcommand{\mcIkb}{{I}_{k,b}}

\newcommand{\Ns}{N_{\tnr{s}}}
\newcommand{\ld}{\ldots}

\newcommand{\bU}[0]{\underline{U}}
\newcommand{\bX}[0]{\underline{X}}\newcommand{\bx}[0]{\underline{x}}
\newcommand{\bL}[0]{\underline{L}}
\newcommand{\bS}[0]{\underline{S}}
\newcommand{\bY}[0]{\underline{Y}}\newcommand{\by}[0]{\underline{y}}

\begin{document}

\title{Information Rates and post-FEC BER Prediction in Optical Fiber Communications}
\author{Alex Alvarado}
\address{Signal Processing Systems (SPS) Group, Department of Electrical Engineering\\
Eindhoven University of Technology (TU/e),5600MB Eindhoven, The Netherlands}
\email{alex.alvarado@ieee.org} 
\begin{abstract}
Information-theoretic metrics to analyze optical fiber communications systems with binary and nonbinary soft-decision FEC are reviewed. The numerical evaluation of these metrics in both simulations and experiments is also discussed. Ready-to-use closed-form approximations are presented.
\end{abstract}
\ocis{(060.4080) Modulation, (060.4510) Optical communications.}

\maketitle

\section{Introduction and Motivation}

Modern fiber optical communication systems are based on multi-level modulation and soft-decision (SD) forward error correction (FEC), a combination known as coded modulation (CM). Ungerboeck's celebrated trellis-coded modulation \cite{Ungerboeck82} was very popular because the receiver could find the most likely coded sequence using a single low-complexity decoder that exploited the CM trellis structure. With the advent of powerful SD-FEC such low-density parity-check (LDPC) codes, however, this CM paradigm has changed. Most modern fiber optical CM transceivers use a receiver that splits the decoding process in two: (i) the received information is converted into \emph{soft symbols} or \emph{soft bits}, and (ii) this soft information is used by a binary or nonbinary SD-FEC. Due to its ease of implementation, the most popular case is when the FEC is binary, which is usually known as bit-interleaved coded modulation \cite{Zehavi92,Alvarado15_Book,Alvarado2015_JLT}. 

Optical transmission experiments usually do not include FEC. This allows new FEC schemes to be designed using recorded data obtained in expensive and time-consuming experiments. In this context, the use of \emph{thresholds} became a simple and powerful way of deciding if the bit error rate (BER) after FEC decoding would be below the BER target (typically around $10^{-15}$). A commonly used threshold in the optical communications literature is the pre-FEC BER, which stemmed from outdated experiments based on \emph{hard-decision} FEC. 

The pre-FEC BER threshold does not work well for nonbinary SD-FEC (NB-SD-FEC) \cite{Schmalen16} nor for binary SD-FEC (B-SD-FEC) \cite{Alvarado2015b_JLT}. Better predictors for the post-FEC BER of SD-FEC are achievable information rates (AIRs) such as the mutual information (MI) and generalized mutual information (GMI). The MI and GMI cannot only be used as accurate decoding thresholds for NB-SD-FEC and B-SD-FEC respectively, but they are also in general better system performance metrics for CM---better than, e.g., the widely used Q-factor---as previously discussed, in e.g., \cite{Secondini13,Fehenberger15a,Eriksson16}.

In this semi-tutorial paper we consider both B-SD-FEC and NB-SD-FEC decoders, whose AIRs were recently compared in \cite{Liga16}. We discuss the advantages of AIRs as a system performance metric as well as methods to numerically evaluate AIRs in both simulations and experiments. Ready-to-use expressions for arbitrary multidimensional constellations in an additive white Gaussian noise (AWGN) channel are presented.

\section{AIRs as a System Performance Metric}

\begin{wrapfigure}{r}{0.54\textwidth}
\vspace{-4ex}
\centerline{
\footnotesize{
\begin{tikzpicture}[>=stealth,auto,
block/.style={rectangle,rounded corners,thick,draw,inner sep=1pt,minimum width=14mm,minimum height=8mm,fill=blue,drop shadow,align=center,execute at begin node=\setlength{\baselineskip}{2ex}},
plain/.style={align=center,execute at begin node=\setlength{\baselineskip}{1.5ex}},
]
\node[plain,inner sep=0] (C) {};
\node[rectangle,align=center,rounded corners,thick,draw,inner sep=1pt,minimum width=1cm,text width=2cm,minimum height=9mm,fill=lightgray,fill opacity=1] at (C) (REC) {Optical Channel $p_{\bY|\bX}(\by|\bx)$}; 
\node[block, left=7mm of REC, draw, thick, fill=orange, rotate=90, anchor=south] (Tx){Mapper};
\node[block, left=4mm of Tx.north, draw, thick, fill=orange, rotate=90, anchor=south] (ENC){Encoder};
\node[block, right=7mm of REC, draw, thick, fill=orange, rotate=90, anchor=north] (Rx){Demapper};
\draw[thick,<-] (ENC) -- ($(-4mm,0)+(ENC.north)$) node[above,inner sep=0.4mm,anchor=south,align=right]{$\underline{U}$};
\node[block, right=4mm of Rx.south, draw, thick, fill=orange, rotate=90, anchor=north] (DEC){Decoder};
\draw[thick,->] (Tx) -- node[above,inner sep=0.4mm]{$\bX$} (REC.west);
\draw[thick,->] (REC.east) -- node[above,inner sep=0.4mm]{$\bY$} (Rx);
\draw[thick,->] (ENC.south) -- node[above,inner sep=0.4mm]{$\bS$} (Tx.north);
\path (ENC.south) -- (Tx.north) node[midway,align=center,inner sep=0] (BL) {};
\draw[thick,->] (Rx.south) -- node[above,inner sep=0.4mm]{$\bL$} (DEC.north);
\node[rectangle,rounded corners,thick,draw,inner sep=0pt,minimum width=23mm,minimum height=17mm] at (BL) (RECL) {}; 
\node[plain,above=0mm of RECL.north west,anchor=south west,inner sep=2pt] (CMEncStr) {CM Encoder}; 
\path (Rx.south) -- (DEC.north) node[midway,align=center,inner sep=0] (BR) {};
\draw[thick,->] (DEC.south) -- ($(+5mm,0)+(DEC.south)$) node[above,inner sep=0.4mm,anchor=south,align=left]{$\hat{\bU}$};
\node[rectangle,rounded corners,thick,draw,inner sep=0pt,minimum width=23mm,minimum height=17mm] at (BR) (RECR) {}; 
\node[plain,above=0mm of RECR.north east,anchor=south east,inner sep=2pt] (CMDecStr) {CM Decoder}; 
\end{tikzpicture}
}}
\vspace{-2ex}
\caption{Coded modulation system under consideration.}
\vspace{-3ex}
\label{model}
\end{wrapfigure}
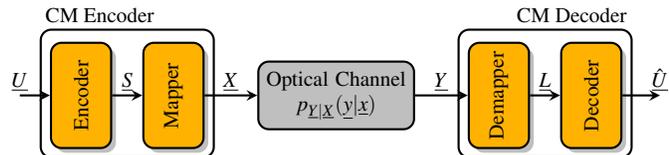
Consider the optical communication system shown in Fig.~\ref{model}. A vector of information bits $\underline{U}$ is mapped to a vector of coded symbols $\underline{X}$, where each symbol is taken from a discrete constellation (e.g, PSK or QAM) with $M$ constellation points. These symbols are then converted into electrical signals, then into optical signals, and then transmitted through a multi-span fiber optical link. The noisy received optical signal is then filtered and converted into an electrical signal, sampled, and digitally processed (including, e.g., equalization, carrier phase estimation, matched filter, etc.). The noisy received symbols $\underline{Y}$ are then converted into soft information $\underline{L}$, which is a measure of how reliable the symbols or bits are. The decoder then obtains an estimate of the information bits $\hat{\underline{U}}$ using this soft information. The FEC encoder/decoder can be binary or nonbinary. When a B-SD-FEC is used, $\underline{S}$ are coded bits and $\underline{L}$ are soft bits. When a NB-SD-FEC is considered, $\underline{S}$ are $M$-ary symbols and $\underline{L}$ are soft symbols.

One of the key advantages of using AIRs as performance metrics is that they are inherently related to FEC. Unlike uncoded metrics like the pre-FEC BER, symbol error rate (SER), EVM, or Q-factor, AIRs give an indication of the amount of \emph{information bits} that can be reliably pushed through a channel. While uncoded metrics are related to bits before and after the demapper, AIRs deal with the information bits before and after FEC. AIRs also allow fair comparisons of different DSP, decoding and nonlinearity compensation techniques, etc., as recently done in \cite{Liga16}.

\begin{figure}[t]
\begin{tikzpicture} 
\begin{semilogyaxis}[
        width=0.5\textwidth,
        height=0.24\textheight,
	every axis/.append style={font=\small},
	xlabel={Normalised MI: $I(X;Y)/\log_2{M}$},
	ylabel= {Post-FEC SER},
        ylabel near ticks,
    	xlabel near ticks,
	ymin=5e-5,ymax=0.5,
	xmin=0.68,xmax=0.98,
	grid=both, 	
]
\addplot [color=blue,mark=square*,mark size=1.5pt, thick,mark options={fill=white,solid}, thick] table [x expr=\thisrow{mi}/3, y={postfec}] {./Figures/paper_figures/nb_sim_qam8_rect_0.70.txt};
\addplot [color=green,mark=square*,mark size=1.5pt, thick,mark options={fill=white,solid}, thick] table [x expr=\thisrow{mi}/3, y={postfec}] {./Figures/paper_figures/nb_sim_qam8_rect_0.75.txt};
\addplot [color=red,mark=square*,mark size=1.5pt, thick,mark options={fill=white,solid}, thick] table [x expr=\thisrow{mi}/3, y={postfec}] {./Figures/paper_figures/nb_sim_qam8_rect_0.80.txt};
\addplot [color=cyan,mark=square*,mark size=1.5pt, thick,mark options={fill=white,solid}, thick] table [x expr=\thisrow{mi}/3, y={postfec}] {./Figures/paper_figures/nb_sim_qam8_rect_0.85.txt};
\addplot [color=magenta,mark=square*,mark size=1.5pt, thick,mark options={fill=white,solid}, thick] table [x expr=\thisrow{mi}/3, y={postfec}] {./Figures/paper_figures/nb_sim_qam8_rect_0.90.txt};
\addplot [color=blue,mark=asterisk,mark size=1.5pt, thick,mark options={fill=white,solid}, thick] table [x expr=\thisrow{mi}/3, y={postfec}] {./Figures/paper_figures/nb_sim_qam8_star_0.70.txt};
\addplot [color=green,mark=asterisk,mark size=1.5pt, thick,mark options={fill=white,solid}, thick] table [x expr=\thisrow{mi}/3, y={postfec}] {./Figures/paper_figures/nb_sim_qam8_star_0.75.txt};
\addplot [color=red,mark=asterisk,mark size=1.5pt, thick,mark options={fill=white,solid}, thick] table [x expr=\thisrow{mi}/3, y={postfec}] {./Figures/paper_figures/nb_sim_qam8_star_0.80.txt};
\addplot [color=cyan,mark=asterisk,mark size=1.5pt, thick,mark options={fill=white,solid}, thick] table [x expr=\thisrow{mi}/3, y={postfec}] {./Figures/paper_figures/nb_sim_qam8_star_0.85.txt};
\addplot [color=magenta,mark=asterisk,mark size=1.5pt, thick,mark options={fill=white,solid}, thick] table [x expr=\thisrow{mi}/3, y={postfec}] {./Figures/paper_figures/nb_sim_qam8_star_0.90.txt};
\addplot [color=blue,mark=triangle*,mark size=1.5pt, thick,mark options={fill=white,solid}, thick] table [x expr=\thisrow{mi}/3, y={postfec}] {./Figures/paper_figures/nb_sim_qam8_sp16_0.70.txt};
\addplot [color=green,mark=triangle*,mark size=1.5pt, thick,mark options={fill=white,solid}, thick] table [x expr=\thisrow{mi}/3, y={postfec}] {./Figures/paper_figures/nb_sim_qam8_sp16_0.75.txt};
\addplot [color=red,mark=triangle*,mark size=1.5pt, thick,mark options={fill=white,solid}, thick] table [x expr=\thisrow{mi}/3, y={postfec}] {./Figures/paper_figures/nb_sim_qam8_sp16_0.80.txt};
\addplot [color=cyan,mark=triangle*,mark size=1.5pt, thick,mark options={fill=white,solid}, thick] table [x expr=\thisrow{mi}/3, y={postfec}] {./Figures/paper_figures/nb_sim_qam8_sp16_0.85.txt};
\addplot [color=magenta,mark=triangle*,mark size=1.5pt, thick,mark options={fill=white,solid}, thick] table [x expr=\thisrow{mi}/3, y={postfec}] {./Figures/paper_figures/nb_sim_qam8_sp16_0.90.txt};
\node at (axis cs:0.76,1e-3) [anchor=north,align=center,rotate=-84,inner sep=0,font=\scriptsize,color=black] {$R_\tnr{c}=0.70$};
\node at (axis cs:0.80,1e-3) [anchor=north,align=center,rotate=-84,inner sep=0,font=\scriptsize,color=black] {$R_\tnr{c}=0.75$};
\node at (axis cs:0.84,1e-3) [anchor=north,align=center,rotate=-84,inner sep=0,font=\scriptsize,color=black] {$R_\tnr{c}=0.80$};
\node at (axis cs:0.88,1e-3) [anchor=north,align=center,rotate=-84,inner sep=0,font=\scriptsize,color=black] {$R_\tnr{c}=0.85$};
\node at (axis cs:0.92,1e-3) [anchor=north,align=center,rotate=-84,inner sep=0,font=\scriptsize,color=black] {$R_\tnr{c}=0.90$};
\end{semilogyaxis}
\end{tikzpicture}
\begin{tikzpicture} 
\begin{semilogyaxis}[
        width=0.5\textwidth,
        height=0.24\textheight,
	every axis/.append style={font=\small},
	xlabel={Normalised GMI: $\textnormal{GMI}/\log_2{M}$},
	ylabel= {Post-FEC BER},
        ylabel near ticks,
    	xlabel near ticks,
	ymin=5e-5,ymax=0.5,
	xmin=0.68,xmax=0.98,
	grid=both, 	
]
\addplot [color=blue,mark=*,mark size=1.5pt, thick,mark options={fill=white,solid}, thick] file {./data/BERout_GMI_4QAM_R_0_71.txt};
\addplot [color=green,mark=*,mark size=1.5pt, thick,mark options={fill=white,solid}, thick] file {./data/BERout_GMI_4QAM_R_3_4.txt};
\addplot [color=red,mark=*,mark size=1.5pt, thick,mark options={fill=white,solid}, thick] file {./data/BERout_GMI_4QAM_R_0_81.txt};
\addplot [color=cyan,mark=*,mark size=1.5pt, thick,mark options={fill=white,solid}, thick] file {./data/BERout_GMI_4QAM_R_0_86.txt};
\addplot [color=magenta,mark=*,mark size=1.5pt, thick,mark options={fill=white,solid}, thick] file {./data/BERout_GMI_4QAM_R_9_10.txt};
\addplot [color=blue,mark=asterisk,mark size=1.5pt, thick,mark options={fill=white,solid}, thick] file {./data/BERout_GMI_16QAM_R_0_71.txt};
\addplot [color=green,mark=asterisk,mark size=1.5pt, thick,mark options={fill=white,solid}, thick] file {./data/BERout_GMI_16QAM_R_3_4.txt};
\addplot [color=red,mark=asterisk,mark size=1.5pt, thick,mark options={fill=white,solid}, thick] file {./data/BERout_GMI_16QAM_R_0_81.txt};
\addplot [color=cyan,mark=asterisk,mark size=1.5pt, thick,mark options={fill=white,solid}, thick] file {./data/BERout_GMI_16QAM_R_0_86.txt};
\addplot [color=magenta,mark=asterisk,mark size=1.5pt, thick,mark options={fill=white,solid}, thick] file {./data/BERout_GMI_16QAM_R_9_10.txt};
\addplot [color=blue,mark=triangle*,mark size=1.5pt, thick,mark options={fill=white,solid}, thick] file {./data/BERout_GMI_64QAM_R_0_71.txt};
\addplot [color=green,mark=triangle*,mark size=1.5pt, thick,mark options={fill=white,solid}, thick] file {./data/BERout_GMI_64QAM_R_3_4.txt};
\addplot [color=red,mark=triangle*,mark size=1.5pt, thick,mark options={fill=white,solid}, thick] file {./data/BERout_GMI_64QAM_R_0_81.txt};
\addplot [color=cyan,mark=triangle*,mark size=1.5pt, thick,mark options={fill=white,solid}, thick] file {./data/BERout_GMI_64QAM_R_0_86.txt};
\addplot [color=magenta,mark=triangle*,mark size=1.5pt, thick,mark options={fill=white,solid}, thick] file {./data/BERout_GMI_64QAM_R_9_10.txt};
\node at (axis cs:0.755,1e-3) [anchor=south,align=center,rotate=-86,inner sep=0,font=\scriptsize,color=black] {$R_\tnr{c}=0.71$};
\node at (axis cs:0.79,1e-3) [anchor=south,align=center,rotate=-86,inner sep=0,font=\scriptsize,color=black] {$R_\tnr{c}=0.75$};
\node at (axis cs:0.842,1e-3) [anchor=south,align=center,rotate=-86,inner sep=0,font=\scriptsize,color=black] {$R_\tnr{c}=0.81$};
\node at (axis cs:0.885,1e-3) [anchor=south,align=center,rotate=-86,inner sep=0,font=\scriptsize,color=black] {$R_\tnr{c}=0.86$};
\node at (axis cs:0.925,1e-3) [anchor=south,align=center,rotate=-86,inner sep=0,font=\scriptsize,color=black] {$R_\tnr{c}=0.90$};
\end{semilogyaxis}
\end{tikzpicture}
\vspace{-1ex}
\caption{Post-FEC SER for NB-LDPC codes in \cite{Schmalen16} (left) and post-FEC BER for (punctured) DVB-S2 LDPC codes (right).}
\vspace{-3ex}
\label{prediction}
\end{figure}
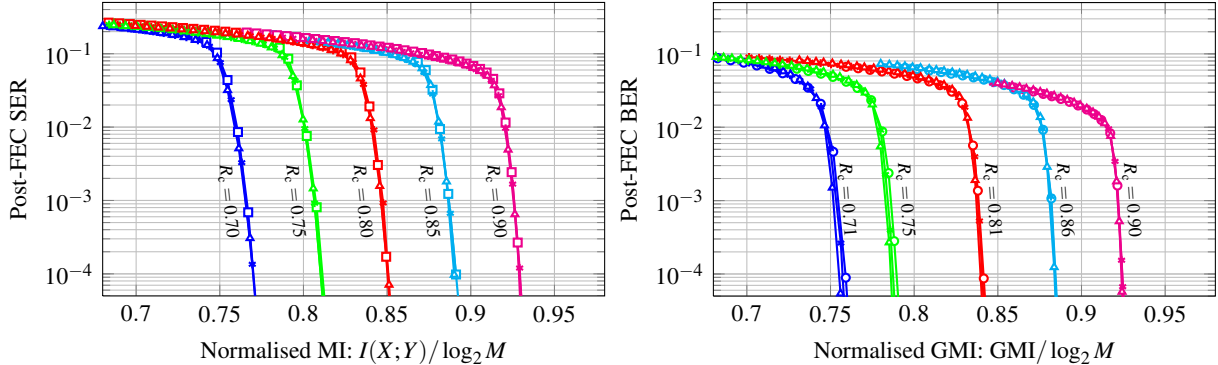

AIRs can also be used as decoding thresholds. The error probability after FEC can be accurately predicted by considering the MI and GMI, for NB-SD-FEC and B-SD-FEC, resp. Consider for example an AWGN channel, the NB-LDPC codes from \cite{Schmalen16} with rates $R_\tnr{c}=\left\{0.70,0.75,0.80,0.85,0.90\right\}$, and three different $8$QAM constellations from \cite[Fig.~3]{Schmalen16}. The post-FEC SER results are shown in Fig.~\ref{prediction} (left), where different markers represent different modulation formats. This figure shows how the MI (normalized by the number of bit/symbol) is a good predictor of the post-FEC SER. Similar results are shown in Fig.~\ref{prediction} (right) for LDPC codes from the DVB-S2 standard and $R_\tnr{c}=\left\{0.71,0.75,0.81,0.86,0.90\right\}$ (obtained by random puncturing of the original codes) and three different modulation formats (4QAM, 16QAM, and 64QAM). In this case the decoder is binary and the normalized GMI is the quantity that correctly predicts the post-FEC BER for all modulation formats (different markers). The main result in Fig.~\ref{prediction} is that normalized MI and GMI are very good decoding thresholds for SD-FEC.

It was shown in \cite{Alvarado2015b_JLT} that the post-FEC BER of B-SD-FEC cannot be predicted using pre-FEC BER. There are multiple reasons for not using pre-FEC BER for SD-FEC. The main one is that that pre-FEC BER is an \emph{uncoded} metric based on \emph{hard decisions} (bits), while CM with SD-FEC are \emph{coded} systems with decoders based on \emph{soft decisions}. A possibly even more compelling reason is that for the NB-SD-FEC case, pre-FEC BER simply cannot be measured, as in this case the receiver operates only on symbols and bits after the demapper are simply not present.

\section{AIRs Computations}

Here we consider arbitrary constellations with variance $\sigma_x^2$ in $N$ complex dimensions. The constellation symbols are transmitted with equal probability over a discrete-time, memoryless AWGN channel. This model represents well the optical channel in Fig.~\ref{model} for uncompensated long-haul systems and when memoryless demappers are used. Under these channel assumptions, the MI can be expressed as the following multidimensional integral
\setlength\abovedisplayskip{0.55ex}
\setlength\belowdisplayskip{0.55ex}
\begin{align}
\label{MD.MI.general.5}
 I(X;Y)       & = \log_2{M}-\frac{1}{M}\sum_{i=1}^{M} \underbrace{\int_{\mathds{C}}\int_{\mathds{C}}\ldots \int_{\mathds{C}}}_{\text{$N$ times}} f_{Z}(z) \log_{2}{\sum_{j=1}^{M}\exp{\left(-\frac{\|\dij\|^2+2\Re\set{\left<z,\dij\right>}}{\sigma_{z}^2/N}\right)}} \, \tnr{d}z,
\end{align}
where $f_{Z}(z)$ is the probability density function of the $N$-dimensional complex random \emph{vector} $Z$ whose elements are circularly-symmetric zero-mean complex Gaussian random variables, $\sigma_{z}^2$ is the total variance of the noise, $\dij=x_i-x_j$, $x_i$ is the $i$-th $N$-dimensional symbol, and $\Re\set{\left<\cdot,\cdot\right>}$ is the real part of the inner product. An expression similar to \eqref{MD.MI.general.5} exists for the GMI, which not only depends on the constellation but also on its binary labeling (bit-to-symbol mapping).

Both MI and GMI for the multidimensional AWGN channel can be efficiently evaluated numerically via Gauss--Hermite quadrature. Closed-form expressions for this were presented in \cite[Sec.~4.5]{Alvarado15_Book}. An alternative (and more general) method to numerically calculate MIs and GMIs is to use Monte Carlo integration to approximate the multidimensional integrals. This method is particularly efficient when the number of dimensions $N$ grows, which for the MI in \eqref{MD.MI.general.5} gives
\begin{align}
\label{MD.MI.MC}
I(X;Y) & \approx \log_2{M} -\frac{1}{M}\sum_{i=1}^{M}\frac{1}{\Ns}\sum_{n=1}^{\Ns} 
\log_{2}{\sum_{j=1}^{M}\exp{
\Biggl(
-\frac{N}{\sigma_{z}^2}\bigl({\|\dij\|^2+2\Re\set{\bigl<z^{(n)},\dij\bigr>}}\bigr)
\Biggr)
}
},
\end{align}
where $z^{(n)}$ with $n=1,2,\ld,\Ns$ are vectors whose elements are independent realizations of circularly-symmetric zero-mean complex Gaussian random variables, each with variance $\sigma_{z}^2/N$. In a simulation or experiment where $M N_s$ symbols were transmitted, the r.h.s. of \eqref{MD.MI.MC} can be estimated in three steps: (i) estimate the noise variance, (ii) for each symbol $x_i$, obtain noise realizations $z^{(n)}$ by substracting the transmitted from the received symbols in all the timeslots where $x_i$ was transmitted, and (iii) use those samples to compute the two innermost sums in \eqref{MD.MI.MC} for all $i=1,\ldots,M$. Although the optical channel is not in general AWGN, this estimated quantity is an AIR using mismatched metrics.

\begin{wrapfigure}{r}{0.48\textwidth}
\vspace{-2ex}
\centering
\includegraphics{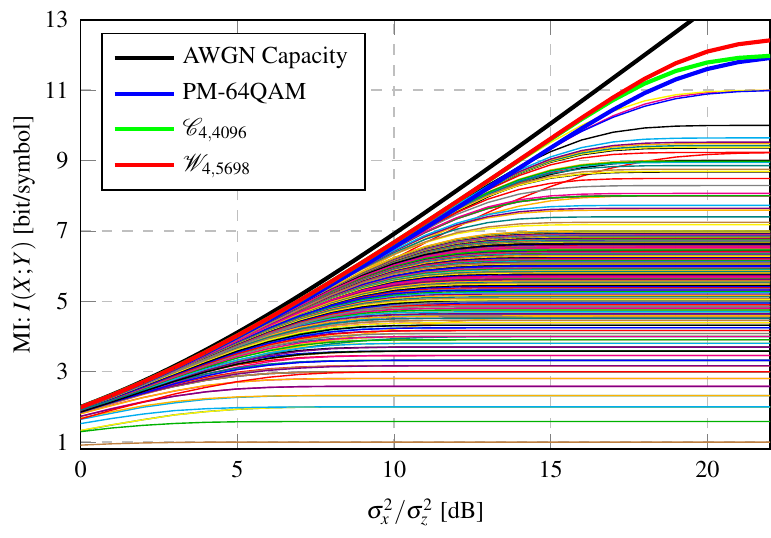}
\vspace{-2ex}
\caption{MI for all 2D complex (4D real) constellations in \cite{Agrell-codes2016}.}
\label{fig:MIs}
\vspace{-3ex}
\end{wrapfigure}
We computed \eqref{MD.MI.MC} for all the $227$ $2$D complex constellations listed in \cite{Agrell-codes2016}. The obtained results are shown in Fig.~\ref{fig:MIs} (colored lines), where the AWGN capacity is shown as reference. All these curves were computed using $N_s=10^4$ Monte Carlo samples and obtained in a few hours on a standard computer. The results in this figure can be used to compare the performance of different modulation formats, even when they have a different number of constellation points. In Fig.~\ref{fig:MIs}, we highlight PM-64QAM ($M=4096$) and two other constellations that outperform PM-64QAM: $\mathcal{C}_{4,4096}$ ($M=4096$), and $\mathcal{W}_{4,5698}$ ($M=5698$), the latter proposed in 1974 in \cite{Welti74}. The constellation $\mathcal{W}_{4,5698}$ shows an excellent performance for a very large range of MIs. We warn the reader, however, to be cautious with MI analysis. Constellations that are good in terms of MI might not be good in terms of GMI, as previously shown in \cite{Alvarado2015_JLT}. This GMI can be approximated via Monte Carlo  as
\begin{align}
\label{MD.GMI.MC}
\tnr{GMI} & \approx \log_2{M}-\frac{1}{M}\sum_{k=1}^{\log_2{M}}\sum_{b\in\set{0,1}}\sum_{i\in\mcIkb} \frac{1}{\Ns} \sum_{n=1}^{\Ns} 
\log_{2}\frac
{\sum_{p=1}^{M}\exp\Bigl(-\frac{N}{\sigma_{z}^2}\bigl({\|\dip\|^{2}+2\Re\set{\bigl<z^{(n)},\dip\bigr>}}\bigr)\Bigr)}
{\sum_{j\in\mcIkb}\exp\Bigl(-\frac{N}{\sigma_{z}^2}\bigl({\|\dij\|^{2}+2\Re\set{\bigl<z^{(n)},\dij\bigr>}}\bigr)\Bigr)},
\end{align}
where $\mcIkb$ is the set of indices of constellation points labeled with a bit $b$ at bit position $k$.

\section{Conclusions}

In this paper, we discussed the usefulness and the numerical calculation of achievable information rates for optical communication systems. The presented expressions are valid for a multidimensional AWGN channel with equally likely symbols. Extensions and generalizations to numerical quadratures, non-Gaussian channels (e.g., for eigenvalue communications), or nonuniform input distributions (for systems based on probabilistic shaping) are straightforward.



\end{document}